\documentclass[aps, showpacs, showkeys,nofootinbib,floatfix]{revtex4}

\usepackage{amssymb}
\usepackage{amsmath}
\usepackage{graphicx}

\begin{document}

\title{Generalized Holographic and Ricci Dark Energy Models}

\author{Lixin Xu\footnote{Corresponding author}}
\email{lxxu@dlut.edu.cn}
\author{Jianbo Lu}
\author{Wenbo Li}

\affiliation{Institute of Theoretical Physics, School of Physics \&
Optoelectronic Technology, Dalian University of Technology, Dalian,
116024, P. R. China}

\begin{abstract}
In this paper, we consider generalized holographic and Ricci dark
energy models where the energy densities are given as
$\rho_{R}=3c^2M^{2}_{pl}Rf(H^2/R)$ and
$\rho_{h}=3c^2M^{2}_{pl}H^2g(R/H^2)$ respectively, here $f(x),g(y)$
are positive defined functions of dimensionless variables $H^2/R$ or
$R/H^2$. It is interesting that holographic and Ricci dark energy
densities are recovered or recovered interchangeably when the
function $f(x)=g(y)\equiv 1$ or$f=g\equiv Id$ is taken respectively
(for example $f(x),g(x)=1-\epsilon(1-x)$, $\epsilon=0 \text{ or } 1$
respectively). Also, when $f(x)\equiv xg(1/x)$ is taken, the Ricci
and holographic dark energy models are equivalents to a generalized
one. When the simple forms $f(x)=1-\epsilon(1-x)$ and
$g(y)=1-\eta(1-y)$ are taken as examples, by using current cosmic
observational data, generalized dark energy models are researched.
As expected, in these cases, the results show that they are
equivalent ($\epsilon=1-\eta=1.312$) and Ricci-like dark energy is
more favored relative to the holographic one where the Hubble
horizon was taken as an IR cut-off. And, the suggestive combination
of holographic and Ricci dark energy components would be $1.312
R-0.312H^2$ which is $2.312H^2+1.312\dot{H}$ in terms of $H^2$ and
$\dot{H}$.
\end{abstract}



\maketitle

\section{Introduction}

The observation of the Supernovae of type Ia
\cite{ref:Riess98,ref:Perlmuter99} provides the evidence that the
universe is undergoing accelerated expansion. Jointing the
observations from Cosmic Background Radiation
\cite{ref:Spergel03,ref:Spergel06} and SDSS
\cite{ref:Tegmark1,ref:Tegmark2}, one concludes that the universe at
present is dominated by $70\%$ exotic component, dubbed dark energy,
which has negative pressure and pushes the universe to accelerated
expansion. To explain the current accelerated expansion, many models
are presented, such as cosmological constant, quintessence
\cite{ref:quintessence1,ref:quintessence2,ref:quintessence3,ref:quintessence4},
phtantom \cite{ref:phantom}, quintom \cite{ref:quintom} and
holographic dark energy \cite{ref:holo1,ref:holo2} etc. For recent
reviews, please see
\cite{ref:DEReview1,ref:DEReview2,ref:DEReview3,ref:DEReview4,ref:DEReview5,ref:DEReview6}.

In particular, a model named holographic dark energy has been
discussed extensively \cite{ref:holo1,ref:holo2,ref:holo3}. The
model is constructed by considering the holographic principle and
some features of quantum gravity theory. According to the
holographic principle, the number of degrees of freedom in a bounded
system should be finite and has relations with the area of its
boundary. By applying the principle to cosmology, one can obtain the
upper bound of the entropy contained in the universe. For a system
with size $L$ and UV cut-off $\Lambda$ without decaying into a black
hole, it is required that the total energy in a region of size $L$
should not exceed the mass of a black hole of the same size, thus
$L^3\rho_{\Lambda} \le L M^2_{pl}$. The largest $L$ allowed is the
one saturating this inequality, thus $\rho_{\Lambda} =3c^2
M^{2}_{pl} L^{-2}$, where $c$ is a numerical constant and $M_{pl}$
is the reduced Planck Mass $M^{2}_{pl}=1/8 \pi G$. It just means a
{\it duality} between UV cut-off and IR cut-off. The UV cut-off is
related to the vacuum energy, and IR cut-off is related to the large
scale of the universe, for example Hubble horizon, event horizon or
particle horizon as discussed by \cite{ref:holo1,ref:holo2}. In the
paper \cite{ref:holo2}, the author takes the future event horizon
\begin{equation}
R_{eh}(a)=a\int^{\infty}_{t}\frac{dt^{'}}{a(t^{'})}=a\int^{\infty}_{a}\frac{da^{'}}{Ha^{'2}}\label{eq:EH}
\end{equation}
as the IR cut-off $L$. This horizon is the boundary of the volume a
fixed observer may eventually observe. One is to formulate a theory
regarding a fixed observer within this horizon. As pointed out in
\cite{ref:holo2}, it can reveal the dynamic nature of the vacuum
energy and provide a solution to the {\it fine tuning} and {\it
cosmic coincidence} problem. In this model, the value of parameter
$c$ determines the property of holographic dark energy. When $c> 1$,
$c=1$ and $c< 1$, the holographic dark energy behaviors like
quintessence, cosmological constant and phantom respectively.
Unfortunately, when the Hubble horizon is taken as the role of IR
cut-off, non-accelerated expansion universe can be achieved
\cite{ref:holo1,ref:holo2,ref:BransDicke}. However, the Hubble
horizon is the most natural cosmological length scale, how to
realize an accelerated expansion by using it as an IR cut-off will
be interesting. One possibility is to generalize the holographic
dark energy model. It will be one of the main points of this work.

Inspired by this principle, Gao, {\it et. al.} took the Ricci scalar
as the IR cut-off and named it Ricci dark energy \cite{ref:Gao},
$\rho_{R}=3c^2M^{2}_{pl}(\dot{H}+2H^2+ k/a^2)\propto R$. In that
paper \cite{ref:Gao}, it has shown that this model can avoid the
causality problem and naturally solve the coincidence problem of
dark energy. Interestingly, Cai, {\it et. al.} found out that the
holographic Ricci dark energy had relations with the causal
connection scale $R^{-2}_{CC}=Max(\dot{H}+2H^2,-\dot{H})$ for a flat
universe \cite{ref:CaiCC}. Also, it was found that only the case
where $R^{-2}_{CC}=\dot{H}+2H^2$ was taken as IR cut-off was
consistent with the current cosmological observations when the
vacuum density appears as an independently conserved energy
component \cite{ref:CaiCC}. The cosmic observational constraints to
the Ricci dark energy model was studied in \cite{ref:HRDConstraint}.
In a manner, one can conclude that $H^2$ or $\dot{H}$ alone can not
provide any late time accelerated expansion of the universe
consistent with cosmic observations. But, their combination will do.
It is the clue that generalized holographic models will be in the
forms of their combinations.

As known, the holographic dark energy and Ricci dark energy both are
candidates of dark energy can explain the late time accelerated
expansion of our universe. It would be interesting to know which one
is the most favored by current cosmic observational data. In
general, we can use the comic observational data as constraints and
implement Bayesian inference and model selection to test the
goodness of models. However, we can test the goodness directly. It
is the byproduct of this work. A generalized model can be designed
to included holographic and Ricci dark energy by introduce a new
parameter which balances holographic and Ricci dark energy model.
The value of the new parameter determines this generalized model
type: holographic, Ricci or a hybrid one. Of course, the best fit
value of the model parameters is determined by cosmic data.

\section{Generalized holographic and Ricci dark energy}\label{sec:GDE}

We consider a Friedmann-Robertson-Walker universe filled with cold
dark matter and dark energy, here it will be holographic dark energy
and Ricci dark energy. Its metric is written as
\begin{equation}
ds^2=-dt^2+a^2(t)\left(\frac{dr^2}{1-kr^2}+r^2d\theta^2+r^2\sin^2\theta
d\phi^2\right),
\end{equation}
where $k=1,0,-1$ for closed, flat and open geometries respectively.
The Friedmann equation is
\begin{equation}
H^2=\frac{8\pi
G}{3}\left(\rho_{m}+\rho_{de}\right)-\frac{k}{a^2},\label{eq:FE}
\end{equation}
where $H$ is the Hubble parameter, $\rho_{m}$ and $\rho_{de}$ denote
the energy densities of cold dark matter and dark energy
respectively.

In \cite{ref:holo1,ref:holo2}, when the Hubble horizon is taken as
the IR cut-off, the holographic dark energy is written as
\begin{equation}
\rho_{h}=3c^2M^{2}_{pl}H^2.
\end{equation}
Unfortunately, it can not give a current accelerated expansion
universe \cite{ref:holo1,ref:holo2,ref:BransDicke} in this case. In
\cite{ref:Gao}, Gao {\it et. al.} suggested the Ricci scalar can be
taken as an IR cut-off, dubbed Ricci dark energy, which is
proportional to the Ricci scalar
\begin{equation}
R=-6\left(\dot{H}+2H^2+\frac{k}{a^2}\right)\label{eq:Ricciscalar}.
\end{equation}
Then, it can be given as
\begin{equation}
\rho_{R}=3c^2M^{2}_{pl}R=3c^2M^{2}_{pl}\left(\dot{H}+2H^2+\frac{
k}{a^2}\right),\label{eq:rhoRDE}
\end{equation}
where {\bf $R$ is the positive part of the Ricci scalar
(\ref{eq:Ricciscalar})} and its coefficient is absorbed in $c^2$.

When the energy components have no interactions and the conservation
equation
\begin{equation}
\dot{\rho}_{de}+3H(1+w_{de})\rho_{de}=0,
\end{equation}
is respected, the equation of state of dark energy $w_{de}$ can be
written as
\begin{equation}
w_{de}=-1-\frac{1}{3}\frac{d\ln\Omega_{de}}{dx}=-1+\frac{(1+z)}{3}\frac{d\ln\Omega_{de}}{dz}.
\end{equation}
where $x=\ln a$, $\Omega_{de}=\rho_{de}/(3M^{2}_{pl}H^2)$ is
dimensionless energy density of dark energy, and the relation
$1/a=1+z$ is used in above equation. The deceleration parameter
$q(z)$ is defined as
\begin{eqnarray}
q&=&-\frac{\ddot{a}}{aH^2}\nonumber\\
&=&-\frac{\dot{H}+H^2}{H^2}\nonumber\\
&=&-1+\frac{(1+z)}{2}\frac{d\ln H^2}{dz}
\end{eqnarray}

In this paper, we will consider a generalized versions of
holographic and Ricci dark energy respectively. They are given in
generalized forms
\begin{eqnarray}
\rho_{GH}&=&3c^2M^{2}_{pl} f\left(\frac{R}{H^2}\right)H^2,\label{eq:GHD}\\
\rho_{GR}&=&3c^2M^{2}_{pl}g\left(\frac{H^2}{R}\right)R,
\label{eq:GRD}
\end{eqnarray}
where $f(x)$ and $g(y)$ are functions of the dimensionless variables
$x=R/H^2$ and $y=H^2/R$ respectively. It is useful to write $R/H^2$
explicitly in the following form in a flat universe
\begin{eqnarray}
\frac{R}{H^2}&=&\frac{\dot{H}+2H^2}{H^2}\nonumber\\
&=&2-\frac{(1+z)}{2}\frac{d\ln H^2}{dz}.
\end{eqnarray}
It can be easily seen that the holographic and Ricci dark energy
models will be recovered when the function $f(x)=g(y)\equiv 1$.
Also, when the function $f(x)=x$ and $g(y)=y$, the holographic and
Ricci dark energy exchange each other. Clearly, the functions can be
written as
\begin{eqnarray}
f\left(\frac{R}{H^2}\right)=1-\epsilon\left(1-\frac{R}{H^2}\right),\label{eq:epsilonGHD}\\
g\left(\frac{H^2}{R}\right)=1-\eta\left(1-\frac{H^2}{R}\right),\label{eq:epsilonGRD}
\end{eqnarray}
where $\epsilon$ and $\eta$ are parameters. In this case, the above
description can be interpreted as follows. When $\epsilon=0$
($\eta=1$) or $\epsilon=1$($\eta=0$), the generalized energy density
becomes holographic(Ricci) and Ricci(holographic) dark energy
density respectively. Also, when the function has the relation
$f(x)=xg(1/x)$ where the variable $x$ is $x=R/H^2$, the holographic
and Ricci dark energy are equivalents to generalized ones. In the
parameterized forms (\ref{eq:epsilonGHD}) and (\ref{eq:epsilonGRD}),
the relation is $\epsilon=1-\eta$. Generally when $\epsilon\neq
0$($\eta\neq 0$) or $\epsilon\neq 1$($\eta\neq 1$), they are hybrid
ones.

In the following sections, we will take Eq. (\ref{eq:epsilonGHD})
and Eq. (\ref{eq:epsilonGRD}) as simple examples to discuss the
properties of the generalized dark energy models in a flat universe
where the model parameters must be determined by cosmic
observations. Once the best fit value of the parameters $\epsilon$
and $\eta$ was found, one can talk about which one is more favored
by cosmic observations. If $\epsilon=0$ ($\eta=1$), the conclusion
holographic dark energy is more favored. Or, one has the opposite
conclusion. In fact, in these two forms, one expects the results
would be that they are equivalent, i.e. $\epsilon=1-\eta$, because
they are the combination in terms $\dot{H}$ and $H^2$. The
parameters just give some balances between these terms. By fitting
cosmic observations, the models' orientation would be found:
holographic- or Ricci-like.

\section{Generalized Holographic and Ricci Dark Energy Models}

\subsection{$f\left(\frac{R}{H^2}\right)=1-\epsilon\left(1-\frac{R}{H^2}\right)$
case}

In this case, the Friedmann equation (\ref{eq:FE}) can be rewritten
as
\begin{eqnarray}
H^2&=&\frac{1}{3M^{2}_{pl}}\left(\rho_{m}+\rho_{GH}\right)\nonumber\\
&=&H^2\Omega_{m}+c^2\left[1-\epsilon\left(1-\frac{R}{H^2}\right)\right]H^2\label{eq:FREGH},
\end{eqnarray}
where $\Omega_{m}=\rho_{m}/(3M^{2}_{pl}H^2)$ is the dimensionless
energy density of dark matter. The corresponding one of generalized
holographic dark energy is
\begin{eqnarray}
\Omega_{GH}&=&\frac{\rho_{GH}}{3M^{2}_{pl}H^2}\nonumber\\
&=&c^2\left[1-\epsilon\left(1-\frac{R}{H^2}\right)\right]\nonumber\\
&=&c^2\left(1+\epsilon-\frac{\epsilon(1+z)}{2}\frac{d \ln
H^2}{dz}\right).\label{eq:OmegaGH}
\end{eqnarray}
The Friedmann Eq. (\ref{eq:FREGH}) can be rewritten as the
differential equation of $H(z)$ with respect to redshift $z$
\begin{equation}
H^2\left\{1-c^2\left[1+\epsilon-\frac{\epsilon(1+z)}{2}\frac{d \ln
H^2}{dz}\right]\right\}=H^2_0\Omega_{m0}(1+z)^3,
\end{equation}
which has the integration
\begin{equation}
H^2(z)=H_0^2\frac{2\Omega_{m0}(1+z)^3+C_0\left[2+c^2(\epsilon-2)\right](1+z)^{2-\frac{2}{c^2\epsilon}+\frac{2}{\epsilon}}}{2+c^2(\epsilon-2)},
\end{equation}
where $C_0$ is an integral constant
\begin{equation}
C_0=\frac{\left[2+c^2(\epsilon-2)\right]-2\Omega_{m0}}{2+c^2(\epsilon-2)}.
\end{equation}

In this case, the deceleration $q(z)$ is
\begin{equation}
q=\frac{1}{\epsilon}-\frac{\Omega_{GH}}{c^2\epsilon}.
\end{equation}

\subsection{$g\left(\frac{H^2}{R}\right)=1-\eta\left(1-\frac{H^2}{R}\right)$
case}

In this case, the Friedmann equation (\ref{eq:FE}) can be rewritten
as
\begin{eqnarray}
H^2&=&\frac{1}{3M^{2}_{pl}}\left(\rho_{m}+\rho_{GR}\right)\nonumber\\
&=&H^2\Omega_{m}+c^2\left[1-\eta\left(1-\frac{H^2}{R}\right)\right]R\label{eq:FREGR}.
\end{eqnarray}
The dimensionless energy density of generalized Ricci dark energy is
\begin{eqnarray}
\Omega_{GR}&=&\frac{\rho_{GR}}{3M^{2}_{pl}H^2}\nonumber\\
&=&c^2\left[1-\eta\left(1-\frac{H^2}{R}\right)\right]\frac{R}{H^2}\nonumber\\
&=&c^2\left[(2-\eta)-(1-\eta)\frac{(1+z)}{2}\frac{d\ln
H^2}{dz}\right].\label{eq:OmegaGR}
\end{eqnarray}
The Friedmann Eq. (\ref{eq:FREGH}) can be rewritten as the
differential equation of $H(z)$ with respect to redshift $z$
\begin{equation}
H^2\left\{1-c^2\left[(2-\eta)-(1-\eta)\frac{(1+z)}{2}\frac{d\ln
H^2}{dz}\right]\right\}=H^2_0\Omega_{m0}(1+z)^3,
\end{equation}
which has the solution
\begin{equation}
H^2(z)=H^2_0\frac{D_0\left[c^2(1+\eta)-2\right](1+z)^{\frac{2}{c^2(\eta-1)}+\frac{2(\eta-2)}{\eta-1}}-2\Omega_{m0}(1+z)^3}{c^2(1+\eta)-2},
\end{equation}
where $C_1$ is an integral constant
\begin{equation}
D_0=\frac{\left[c^2(1+\eta)-2\right]+2\Omega_{m0}}{c^2(1+\eta)-2}.
\end{equation}
The deceleration parameter is
\begin{equation}
q=\frac{1}{1-\eta}-\frac{\Omega_{GR}}{(1-\eta)c^2}.
\end{equation}

One can immediately find out that they are equivalent when
\begin{equation}
\epsilon=1-\eta
\end{equation}
from the comparison of Eq. (\ref{eq:OmegaGH}) and Eq.
(\ref{eq:OmegaGR}). That can also be seen from the expression of
deceleration parameter $q$.

\subsection{Comparison and Discussion}

It would be interesting to investigate how similar are the
generalized holographic and Ricci dark energy models when the
parameters are given by current cosmic observations. In the other
words, we are going to understand which one is the most favored by
confronting the cosmic observations. As shown in above subsections,
they are equivalent when $\epsilon=1-\eta$ is respected. So, we can
take one of them to investigate its properties. Here, we take the
generalized holographic dark energy model as an example, the
corresponding results of generalized Ricci one can be obtained by
replacing $\eta=1-\epsilon$.

\begin{figure}[tbh]
\centering
\includegraphics[width=5.0in]{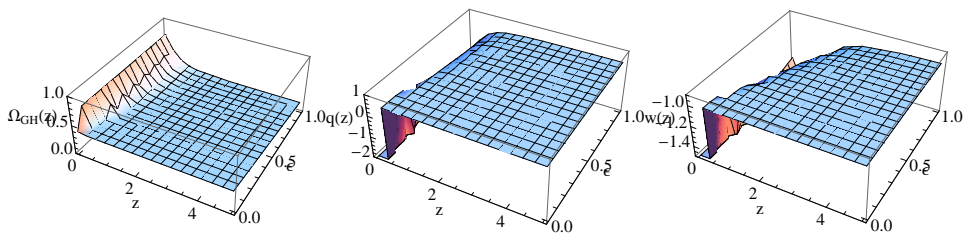}
\caption{The 3D plots of $\Omega_{GH}(z)$, $q(z)$ and $w(z)$ in the
case of generalized holographic dark energy models with redshift $z$
and $\epsilon$ where the values of parameters $c=0.6$, $H_0=72$ and
$\Omega_{m0}=0.27$ are adopted. The corresponding results of
generalized Ricci dark energy model can be obtained by reflection of
the plane $\epsilon=1/2$.}\label{fig:3dH}
\end{figure}

In Fig. \ref{fig:3dH}, the 3D plots of $\Omega(z)$, $q(z)$ and
$w(z)$ respectively with respect redshift $z$ and $\epsilon$ are
presented in generalized holographic dark energy model where the
value of parameter $c=0.6$, $\Omega_{m0}=0.27$ is adopted. From the
figures, one can see that, in the generalized form of holographic
dark energy model, a late time accelerated expansion of our universe
is realized. That can not be obtained in holographic dark energy
model where the Hubble horizon is taken as an IR cut-off. The reason
may be that the Ricci component fills the missing gap or remedies
the fake. By a further investigation, one would find out that the
term $\dot{H}$ has the main effect. Here, we take this term with
$H^2$, i.e. the Ricci scalar, as a whole. Also, one can find the
properties of the generalized holographic dark energy is also
determined by the parameter $\epsilon$ besides the parameter $c$.
When $c$ and $z$ are fixed, with $\epsilon$ increasing in late time
($z<1$), the transition redshift from decelerated expansion to
accelerated expansion is also increased, but decreasing of the EoS
$w$ and $\Omega_{GH}$ of the generalized holographic dark energy.
One would notice that in these plots the boundary values of
parameter $\epsilon$ (i.e. $\epsilon=0$, or $\epsilon=1$ is not
included in figures.). Also, one can find the corresponding results
of generalized Ricci dark energy model by reflection of the plane
$\epsilon=1/2$.

Now, it is proper to present the constraint results by using cosmic
observations: SN Ia, BAO and CMB shift parameter $R$, for the
details please see Appendix \ref{sec:append}. After calculation, the
results are listed in Tab. \ref{tab:result}.
\begin{table}[tbh]
\begin{center}
\begin{tabular}{c|ccccc}
\hline\hline
Models & $\chi^2_{min}$ & $\Omega_{m0}(1\sigma)$ & $c(1\sigma)$ & $\epsilon(1\sigma)$ & $\chi^2_{min}/dof$\\
\hline GH & $316.855$  & $0.325^{+0.039}_{-0.035}$  & $0.579^{+0.030}_{-0.029}$ & $1.312^{+0.353}_{-0.293}$  & $1.035$ \\
\hline
\end{tabular}
\caption{The minimum values of $\chi^2$ and best fit values of the
parameters of generalized holographic dark energy models. The
corresponding results of generalized Ricci dark energy model can be
obtained by reflection of the plane $\epsilon=1/2$. Here $dof$
denotes the model degrees of freedom.}\label{tab:result}
\end{center}
\end{table}

From the best fit value of $\epsilon$ in Tab. \ref{tab:result}, one
can conclude that the generalized holographic and Ricci dark energy
both incline to the Ricci side in the $\epsilon$ axis
($\epsilon\rightarrow 1$ in GH model, and $\epsilon\rightarrow 0$ in
GR model) relative to the holographic side. Also, the interval to
Ricci dark energy point is about $0.312$. It means the cosmic data
favor a generalized dark energy model which is more Ricci-like. And,
the suggestive combination of holographic and Ricci dark energy
components would be $1.312 R-0.312H^2$ which is
$2.312H^2+1.312\dot{H}$ in terms of $H^2$ and $\dot{H}$.

\begin{figure}[tbh]
\centering
\includegraphics[width=3.0in]{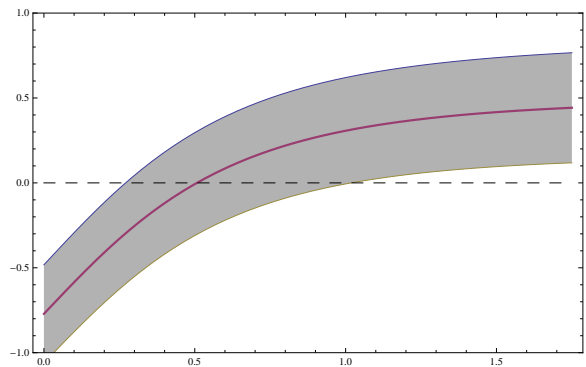}
\caption{The evolution curve of $q(z)$ with redshift $z$ associated
with $1\sigma$ region in the case of generalized holographic dark
energy models where the best fit values of model parameters are
adopted.}\label{fig:qz}
\end{figure}

\begin{figure}[tbh]
\centering
\includegraphics[width=5.0in]{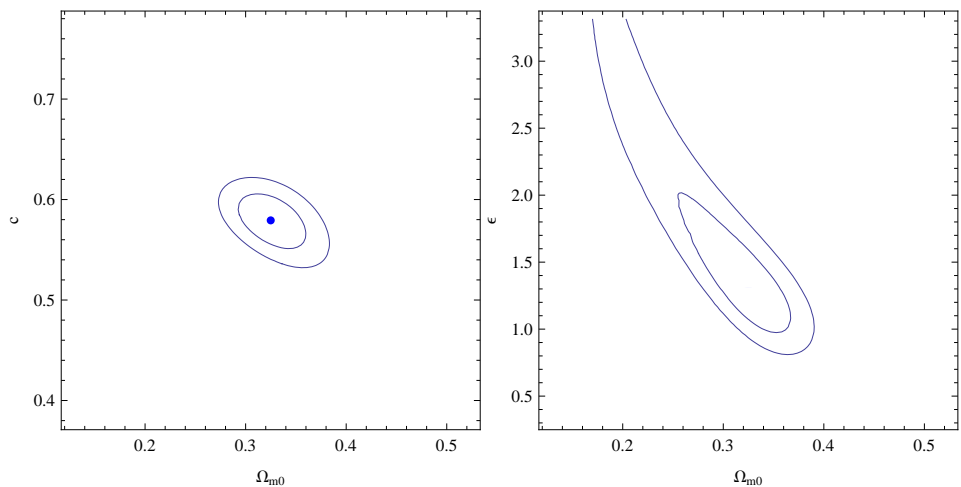}
\caption{The contours in the planes of $\Omega_{m0}-c$ and
$\Omega_{m0}-\epsilon$ with $1\sigma$ and $2\sigma$ regions. The
dots denote the best fit values of model parameters. The
corresponding results of generalized Ricci dark energy model can be
obtained by reflection of the plane $\epsilon=1/2$.}\label{fig:cons}
\end{figure}

The evolution curve of $q(z)$ with respect to redshift $z$ is
plotted in Fig. \ref{fig:qz}. It is clear that, with these best fit
values of model parameters, an late time accelerated expansion of
our universe is obtained. The corresponding contour plots of model
parameters can be found in Fig. \ref{fig:cons}. The transition
redshift from decelerated expansion to accelerated expansion
$z_t=0.507^{+0.512}_{-0.236}$ with $1\sigma$ region is found.

\section{Conclusion}

In this paper, generalized holographic and Ricci dark energy models
are presented, where the energy densities are given as
$\rho_{R}=3c^2M^{2}_{pl}Rf(H^2/R)$ and
$\rho_{h}=3c^2M^{2}_{pl}H^2g(R/H^2)$ respectively, here $f(x),g(y)$
are positive defined functions of dimensionless variables $H^2/R$ or
$R/H^2$. With these generalized forms, the holographic and Ricci
dark energy densities are recovered or recovered interchangeably
when the function $f(x)=g(y)\equiv 1$ or $f=g\equiv Id$ is taken
respectively. As simple examples, we assume the forms of functions
as $f(x)=1-\epsilon(1-x)$ and $g(y)=1-\eta(1-y)$. In these simple
forms, one can immediately find that they are equivalent when
$\epsilon=1-\eta$. It means the results of generalized holographic
and Ricci dark energy are symmetric by reflection of the plane
$\epsilon=1/2$. The best fit values of model parameters are obtained
by using current cosmic observational data as constraints. The
results show that an accelerated expansion of our universe can be
obtained in generalized holographic dark energy model with contrast
to holographic dark energy model where the Hubble horizon is taken
as an IR cut-off. The generalized holographic and Ricci dark energy
both incline to the Ricci side in the $\epsilon$ axis
($\epsilon\rightarrow 1$ in GH model, and $\eta\rightarrow 0$ the
$\eta$ axis in GR model) relative to the holographic side. And, the
interval to Ricci dark energy point is about $0.312$. It means the
cosmic data favor a generalized dark energy model which is more
Ricci-like. And, the suggestive combination of holographic and Ricci
dark energy components would be $1.312 R-0.312H^2$ which is
$2.312H^2+1.312\dot{H}$ in terms of $H^2$ and $\dot{H}$. Of course,
in phenomenological level, one can assume other forms of the
generalized functions $f(x)$ and $g(y)$ to explore the possible
properties of dark energy. We expect this kind of investigation can
shed light on the research of dark energy.

\acknowledgements{This work is supported by NSF (10703001), SRFDP
(20070141034) of P.R. China.}

\appendix

\section{Cosmic Observations}\label{sec:append}

\subsection{SN Ia}
We constrain the parameters with the Supernovae Cosmology Project
(SCP) Union sample including $307$ SN Ia \cite{ref:SCP}, which
distributed over the redshift interval $0.015\le z\le 1.551$.
Constraints from SN Ia can be obtained by fitting the distance
modulus $\mu(z)$
\begin{equation}
\mu_{th}(z)=5\log_{10}(D_{L}(z))+\mu_{0},
\end{equation}
where, $D_{L}(z)$ is the Hubble free luminosity distance $H_0
d_L(z)/c$ and
\begin{eqnarray}
d_L(z)&=&c(1+z)\int_{0}^{z}\frac{dz^{\prime}}{H(z^{\prime})}\\
\mu_0&\equiv&42.38-5\log_{10}h,
\end{eqnarray}
where $H_0$ is the Hubble constant which is denoted in a
re-normalized quantity $h$ defined as $H_0 =100 h~{\rm km ~s}^{-1}
{\rm Mpc}^{-1}$. The observed distance moduli $\mu_{obs}(z_i)$ of SN
Ia at $z_i$ is
\begin{equation}
\mu_{obs}(z_i) = m_{obs}(z_i)-M,
\end{equation}
where $M$ is their absolute magnitudes.

For SN Ia dataset, the best fit values of parameters in a model can
be determined by the likelihood analysis is based on the calculation
of
\begin{eqnarray}
\chi^2(p_s,m_0)&\equiv& \sum_{SNIa}\frac{\left[
\mu_{obs}(z_i)-\mu_{th}(p_s,z_i)\right]^2} {\sigma_i^2} \nonumber\\
&=&\sum_{SNIa}\frac{\left[ 5 \log_{10}(D_L(p_s,z_i)) - m_{obs}(z_i)
+ m_0 \right]^2} {\sigma_i^2}, \label{chi2}
\end{eqnarray}
where $m_0\equiv\mu_0+M$ is a nuisance parameter (containing the
absolute magnitude and $H_0$) that we analytically marginalize over
\cite{ref:SNchi2},
\begin{equation}
\tilde{\chi}^2(p_s) = -2 \ln \int_{-\infty}^{+\infty}\exp \left[
-\frac{1}{2} \chi^2(p_s,m_0) \right] dm_0 \; ,
\label{chi2_marginalization}
\end{equation}
to obtain
\begin{equation}
\tilde{\chi}^2 =  A - \frac{B^2}{C} + \ln \left(
\frac{C}{2\pi}\right) , \label{chi2_marginalized}
\end{equation}
where
\begin{equation}
A=\sum_{SNIa} \frac {\left[5\log_{10}
(D_L(p_s,z_i))-m_{obs}(z_i)\right]^2}{\sigma_i^2},
\end{equation}
\begin{equation}
B=\sum_{SNIa} \frac {5
\log_{10}(D_L(p_s,z_i)-m_{obs}(z_i)}{\sigma_i^2},
\end{equation}
\begin{equation}
C=\sum_{SNIa} \frac {1}{\sigma_i^2} \; .
\end{equation}
The Eq. (\ref{chi2}) has a minimum at the nuisance parameter value
$m_0=B/C$. Sometimes, the expression
\begin{equation}
\chi^2_{SNIa}(p_s,B/C)=A-(B^2/C)\label{eq:chi2SNIa}
\end{equation}
is used instead of Eq. (\ref{chi2_marginalized}) to perform the
likelihood analysis. They are equivalent, when the prior for $m_0$
is flat, as is implied in (\ref{chi2_marginalization}), and the
errors $\sigma_i$ are model independent, what also is the case here.

To determine the best fit parameters for each model, we minimize
$\chi^2(p_s,B/C)$ which is equivalent to maximizing the likelihood
\begin{equation}
{\cal{L}}(p_s) \propto e^{-\chi^2(p_s,B/C)/2} .
\end{equation}

\subsection{BAO}
The BAO are detected in the clustering of the combined 2dFGRS and
SDSS main galaxy samples, and measure the distance-redshift relation
at $z = 0.2$. BAO in the clustering of the SDSS luminous red
galaxies measure the distance-redshift relation at $z = 0.35$. The
observed scale of the BAO calculated from these samples and from the
combined sample are jointly analyzed using estimates of the
correlated errors, to constrain the form of the distance measure
$D_V(z)$ \cite{ref:Okumura2007,ref:Eisenstein05,ref:Percival}
\begin{equation}
D_V(z)=\left[(1+z)^2 D^2_A(z) \frac{cz}{H(z)}\right]^{1/3},
\label{eq:DV}
\end{equation}
where $D_A(z)$ is the proper (not comoving) angular diameter
distance which has the following relation with $d_{L}(z)$
\begin{equation}
D_A(z)=\frac{d_{L}(z)}{(1+z)^2}.
\end{equation}
Matching the BAO to have the same measured scale at all redshifts
then gives \cite{ref:Percival}
\begin{equation}
D_{V}(0.35)/D_{V}(0.2)=1.812\pm0.060.
\end{equation}
Then, the $\chi^2_{BAO}(p_s)$ is given as
\begin{equation}
\chi^2_{BAO}(p_s)=\frac{\left[D_{V}(0.35)/D_{V}(0.2)-1.812\right]^2}{0.060^2}\label{eq:chi2BAO}.
\end{equation}

\subsection{CMB shift Parameter R}

The CMB shift parameter $R$ is given by \cite{ref:Bond1997}
\begin{equation}
R(z_{\ast})=\sqrt{\Omega_m H^2_0}(1+z_{\ast})D_A(z_{\ast})/c
\end{equation}
which is related to the second distance ratio
$D_A(z_\ast)H(z_\ast)/c$ by a factor $\sqrt{1+z_{\ast}}$. Here the
redshift $z_{\ast}$ (the decoupling epoch of photons) is obtained by
using the fitting function \cite{Hu:1995uz}
\begin{equation}
z_{\ast}=1048\left[1+0.00124(\Omega_bh^2)^{-0.738}\right]\left[1+g_1(\Omega_m
h^2)^{g_2}\right],
\end{equation}
where the functions $g_1$ and $g_2$ are given as
\begin{eqnarray}
g_1&=&0.0783(\Omega_bh^2)^{-0.238}\left(1+ 39.5(\Omega_bh^2)^{0.763}\right)^{-1}, \\
g_2&=&0.560\left(1+ 21.1(\Omega_bh^2)^{1.81}\right)^{-1}.
\end{eqnarray}
The 5-year {\it WMAP} data of $R(z_{\ast})=1.710\pm0.019$
\cite{ref:Komatsu2008} will be used as constraint from CMB, then the
$\chi^2_{CMB}(p_s)$ is given as
\begin{equation}
\chi^2_{CMB}(p_s)=\frac{(R(z_{\ast})-1.710)^2}{0.019^2}\label{eq:chi2CMB}.
\end{equation}

For Gaussian distributed measurements, the likelihood function
$L\propto e^{-\chi^2/2}$, where $\chi^2$ is
\begin{equation}
\chi^2=\chi^2_{SNIa}+\chi^2_{BAO}+\chi^2_{CMB},
\end{equation}
where $\chi^2_{SNIa}$ is given in Eq. (\ref{eq:chi2SNIa}),
$\chi^2_{BAO}$ is given in Eq. (\ref{eq:chi2BAO}), $\chi^2_{CMB}$ is
given in Eq. (\ref{eq:chi2CMB}). In this paper, the central values
of $\Omega_b h^2=0.02265\pm 0.00059$, $\Omega_m h^2=0.1369\pm
0.0037$ from 5-year {\it WMAP} results \cite{ref:Komatsu2008} and
$H_0=72\pm{\rm 8 km s^{-1}Mpc^{-1}}$ are adopted.

\end{document}